\documentclass[aps,pra,preprint,amsmath,amssymb,superscriptaddress]{revtex4}
\usepackage{graphicx,epsfig,epsf}
\usepackage{dcolumn}
\usepackage{bm}

\begin{document}
\newcommand{\ri}{ i}
\newcommand{\re}{ e}
\newcommand{\bx}{{\mathbf x}}
\newcommand{\bd}{{\mathbf d}}
\newcommand{\be}{{\mathbf e}}
\newcommand{\br}{{\mathbf r}}
\newcommand{\bk}{{\mathbf k}}
\newcommand{\bE}{{\mathbf E}}
\newcommand{\bI}{{\mathbf I}}
\newcommand{\bR}{{\mathbf R}}
\newcommand{\bZero}{{\mathbf 0}}
\newcommand{\bM}{{\mathbf M}}
\newcommand{\bX}{{\mathbf X}}
\newcommand{\bn}{{\mathbf n}}
\newcommand{\bs}{{\mathbf s}}
\newcommand{\tbs}{\tilde{\mathbf s}}
\newcommand{\rSi}{{\rm Si}}
\newcommand{\beps}{\mbox{\boldmath{$\epsilon$}}}
\newcommand{\bGamma}{\mbox{\boldmath{$\Gamma$}}}
\newcommand{\rg}{{\rm g}}
\newcommand{\tr}{{\rm tr}}
\newcommand{\xmax}{x_{\rm max}}
\newcommand{\xb}{\overline{x}}
\newcommand{\pb}{\overline{p}}
\newcommand{\ra}{{\rm a}}
\newcommand{\rx}{{\rm x}}
\newcommand{\rs}{{\rm s}}
\newcommand{\rP}{{\rm P}}
\newcommand{\up}{\uparrow}
\newcommand{\down}{\downarrow}
\newcommand{\hc}{H_{\rm cond}}
\newcommand{\kb}{k_{\rm B}}
\newcommand{\cI}{{\cal I}}
\newcommand{\tit}{\tilde{t}}
\newcommand{\cE}{{\cal E}}
\newcommand{\cC}{{\cal C}}
\newcommand{\Ubs}{U_{\rm BS}}
\newcommand{\sech}{{\rm sech}}
\newcommand{\qq}{{\mathbf ???}}
\newcommand*{\etal}{\textit{et al.}}
\def\vec#1{\mathbf{#1}}
\def\ket#1{|#1\rangle}
\def\bra#1{\langle#1|}
\def\keps{\mathbf{k}\boldsymbol{\varepsilon}}
\def\dm{\boldsymbol{\wp}}


\title{Precision measurements with photon-subtracted or photon-added Gaussian states}
\author{Daniel Braun}
\affiliation{Universit\'e de Toulouse and CNRS; Laboratoire de
 Physique Th\'eorique (IRSAMC); F-31062 Toulouse, France}
\affiliation{Institut f\"ur theoretische Physik, Universit\"at T\"ubingen,
72076 T\"ubingen}
\author{Pu Jian}
\affiliation{Laboratoire Kastler Brossel, Universit\'e Pierre et Marie Curie--Paris 6, ENS, CNRS; 4 place Jussieu, 75252 Paris, France}
\author{Olivier Pinel}
\affiliation{Centre for Quantum Computation and Communication Technology, Department of Quantum Science, \\
The Australian National University, Canberra, ACT 0200, Australia}
\author{Nicolas Treps}
\affiliation{Laboratoire Kastler Brossel, Universit\'e Pierre et Marie Curie--Paris 6, ENS, CNRS; 4 place Jussieu, 75252 Paris, France}

\centerline{\today}
\begin{abstract}
Photon-subtracted and photon-added Gaussian states are amongst the simplest
non-Gaussian states that are experimentally available. It is generally believed that they are some of the best candidates to enhance sensitivity in parameter extraction. We derive here the quantum Cram\'er-Rao bound for such states and find that 
  for
 large photon numbers photon-subtraction or -addition only leads to a small
 correction of the 
 quantum  Fisher  information (QFI). On the other hand a divergence of the QFI appears for very
 small squeezing in the
 limit of vanishing photon number in the case of photon subtraction, implying
 an arbitrarily precise measurement with almost no light.  However, at
 least for 
 the standard and experimentally established preparation scheme, the
 decreasing success probability of the preparation in that limit exactly
 cancels the divergence, leading to finite sensitivity per square root of
 Hertz, when the duration of the preparation is taken into account. 
\end{abstract}
\maketitle

\section{Introduction}
Propelled by the perspective of quantum information applications, the
creation and use of non-classical states of light  has seen a large increase
of interest in recent years. Classical light is often understood as light
which allows a description through a well-defined positive $P$-function,
whereas all other states are called non-classical
\cite{Kim05,Mandel86}. 
But different classes of non-classical quantum
states can be 
considered. 

A first category are squeezed Gaussian
states \cite{Hudson74,Soto83,PhysRevA.79.062302}. Gaussian states
include a large variety of experimentally relevant states 
that can be produced with high photon numbers, such as coherent states,
single- or multi-mode squeezed states (and therefore entangled states), and
thermal states. Combining squeezed vacuum with a bright coherent state on a
beam-splitter is an important experimental tool for achieving highly
sensitive measurements \cite{Caves81,Keller08,abadie_gravitational_2011}. 

As second category
which is relevant for quantum information is the class
of states whose Wigner function is  
not positive everywhere. For brevity we call such states negative
Wigner function states.  
This type of non-classicality is
motivated amongst other things by the 
continuous variable version of the Gottesmann-Knill theorem which states
that universal quantum computation with continuous variables requires
Hamiltonians that are at least cubic in the quadrature operators
\cite{braunstein_quantum_2005}. Moreover, it was recently shown that quantum algorithms for which
both the initial state and the following operations can be represented by
positive Wigner functions can be simulated efficiently on a classical
computer \cite{mari_positive_2012}.

One of the most promising approaches to negative Wigner function states of light is
through photon subtraction or addition.  Proposed theoretically at the end
of last century \cite{dakna_generating_1997}, photon subtraction from a
coherent state of light was 
realized experimentally by Grangier et al.~in
2004 \cite{wenger_non-gaussian_2004}. Since then a multitude of extensions
have been found or proposed, including photon addition,
multiple photon subtraction, coherent superposition of addition and
subtraction of photons, or subtraction from more general states of light
(see \cite{kim_recent_2008,neergaard-nielsen_photon_2011} and references therein). This kind of
non-classical light can be used for increasing entanglement and consequently efficiency of quantum
teleportation protocols, for the demonstration of non-locality and loophole-free
violation of 
Bell's inequalities, for generating Schr\"odinger kitten states, for quantum 
computing, for noise-less probabilistic amplification, and for the experimental
verification of the bosonic commutation relations
\cite{ourjoumtsev_increasing_2007,lee_enhancing_2011,navarrete-benlloch_enhancing_2012,ferreyrol_implementation_2010,xiang_heralded_2010,bimbard_quantum-optical_2010,zavatta_experimental_2009,takahashi_generation_2008,gerrits_generation_2010,nha_proposed_2004,garcia-patron_proposal_2004,dakna_generating_1997,parigi_probing_2007,bartley_strategies_2013,lee_increasing_2013,kim_quantum_2012,kim_recent_2008,neergaard-nielsen_photon_2011}.

The ultimate sensitivity limits of quantum parameter estimation with
Gaussian states, both 
pure and mixed,  and the multi-parameter limits for the single-mode case, are now fully understood
\cite{Pinel12,Pinel13,monras_phase_2013}. As long as the parameter to
be estimated does not 
depend on the number of photon itself, Gaussian states always lead to a
best scaling as $1/\sqrt{N}$ with the average photon number $N$. However, the
prefactor depends on the squeezing. In \cite{Pinel12} the optimal
measurement strategy was 
identified for finite squeezing resources which makes use of a
specific detection mode.  

So far it was unknown if the relatively
simple procedure of subtracting (or adding)  photons from (or to)
Gaussian states can substantially enhance the sensitivity with which certain
parameters coded in the state of light can be measured, and in particular if
it is possible to beat the standard quantum limit (SQL) this way.  The
latter corresponds to the sensitivity achievable with a coherent state and
is characterized by  a $1/\sqrt{N}$ scaling of the sensitivity with the mean
photon number $N$ (see e.g.~\cite{Giovannetti04}). 

In this paper we provide an answer to this question by 
calculating the 
quantum Cram\'er-Rao bound for the sensitivity with which a parameter 
characterizing the original Gaussian state can be measured after addition or
subtraction of a photon. We show that for
 large photon numbers $N$ single photon-subtraction or -addition only leads to
 a  correction of order $1/N$ of the 
 quantum  Fisher  information (QFI).  Surprisingly, however, a divergence of
 the QFI appears for very 
 small squeezing in the
 limit of vanishing photon number in the case of photon subtraction, implying
 an arbitrary precise measurement with almost no light.  However, at least for
 the standard and experimentally established preparation scheme, the
 decreasing success probability of the preparation in that limit exactly
 cancels the divergence, leading to finite sensitivity per square root of
 Hertz, when the duration of the preparation is taken into
 account. Nevertheless, these results may find application in niches where
 precise measurements are required with almost no light, as for example in the
 context of biological samples \cite{wolfgramm_entanglement-enhanced_2012}.

\section{Photon subtraction and addition}
Different approaches to photon-subtracted states are found in the
literature. Kim et al. \cite{Kim05} consider a physical
process of photon-subtraction close to the experimental procedure
\cite{wenger_non-gaussian_2004}, where the original state 
consists of squeezed vacuum that 
passes a beam splitter. Single photon detection is implemented in one
output mode, and the detection of a single photon in that mode heralds a
photon-subtracted state in the other output mode, whose properties can be
verified with standard Wigner-function reconstruction techniques.  This
approach allowed Kim et al.~to take into account losses and analyze
conditions for observing the 
negativity of the Wigner function. 

We take a simpler approach, following \cite{kim_quantum_2012}, and define a photon-subtracted state relative to a reference state $\hat{\rho}$ in the single mode
case  by  
\begin{equation} \label{rhom}
\hat{\rho}^{-}=\hat{a}\hat{\rho} \hat{a}^\dagger/N^-\,
\end{equation}
where $\hat{a}$ is the photon annihilation operator, $\hat{a}^\dagger$ is the photon creation operator, and $N^-=\tr(\hat{a}^\dagger \hat{a} \hat{\rho})=N$ is the mean photon number.
A photon-added state is defined correspondingly as 
\begin{equation} \label{rhop}
\hat{\rho}^+=\hat{a}^\dagger\hat{\rho} \hat{a}/N^+\,,
\end{equation}
with $N^+=\tr(\hat{a} \hat{a}^\dagger\hat{\rho})=N+1$. This definition immediately implies
that a single coherent state 
$|\alpha\rangle$ is invariant under photon subtraction, but not
under photon addition.  
We will come back to the question of state preparation and its
impact on the experimentally relevant sensitivities in Sec.~\ref{sec.spe}

The Wigner function of a single mode state $\hat{\rho}$ as function of the quadratures
$x$ and $p$ is defined as \cite{GardinerZoller04,Schleich01}
\begin{equation} \label{}
W(x,p)[\hat{\rho}]=\frac{1}{\pi}\int dy \langle x+y|\hat{\rho}|x-y\rangle e^{-2iyp}\,.
\end{equation}
Here and in the following we set $\hbar=1$, such that $\hat{x}=(\hat{a}+\hat{a}^\dagger)/\sqrt{2}$ and $\hat{p}=i(\hat{p}-\hat{x})/\sqrt{2}$. States $\hat{x}\hat{\rho}$,
$\hat{p}\hat{\rho}$, etc., then have Wigner functions (see 
Eq.~(4.5.11) in \cite{GardinerZoller04})
\begin{eqnarray}
W(x,p)[\hat{x}\hat{\rho}]&=&\left(x+\frac{i}{2}\partial_p\right)W(x,p)[\hat{\rho}]\label{xr}\\
W(x,p)[\hat{\rho}\hat{x}]&=&\left(x-\frac{i}{2}\partial_p\right)W(x,p)[\hat{\rho}]\label{rx}\\
W(x,p)[\hat{p}\hat{\rho}]&=&\left(p-\frac{i}{2}\partial_x\right)W(x,p)[\hat{\rho}]\label{pr}\\
W(x,p)[\hat{\rho}\hat{p}]&=&\left(p+\frac{i}{2}\partial_x\right)W(x,p)[\hat{\rho}]\,.\label{rp}
\end{eqnarray}
The general $M$-mode case is obtained in a
completely analogous fashion by simply replacing the single quadratures
$x,p$ with a vector $\bX^t=(x_1,p_1,\ldots,x_M,p_M)$, and adding a label for
mode $k$ in which the photon 
is added/subtracted. We
write the corresponding density matrix as $\hat{\rho}^{(\pm,k)}$. Combining the
definitions of $\hat{a}$, $\hat{a}^\dagger$ with Eq.~(\ref{rhom}), Eq.~(\ref{rhop}), and
Eqs.~(\ref{xr}-\ref{rp}), we find
for the
Wigner function
\begin{equation} \label{Wpm}
W(\bX)[\hat{\rho}^{(\pm,k)}]=\frac{1}{2}\left(x_k^2+p_k^2\mp x_k\partial_{x_k}\mp
p_k\partial_{p_k}+\frac{1}{4}(\partial_{p_k}^2+\partial_{x_k}^2)+1\right)W(\bX)[\hat{\rho}]/N^\pm\,. 
\end{equation}

\section{Quantum Cram\'er-Rao bound}
Quantum parameter estimation theory (QPET) establishes the ultimate lower bound to
the sensitivity with which a classical parameter $\theta$ that parametrizes
the quantum state can be measured.  This sensitivity is fundamentally due to
quantum fluctuations, and becomes relevant once all other sources of noise,
error and imperfection are eliminated. QPET generalizes classical parameter
estimation theory (PET), which sets a lower bound on the fluctuations with which a
parameter characterizing a probability distribution $p( \theta,A)$ of
measurement outcomes $A_i$ of an observable $A$ can be estimated. It is
optimized over all possible estimator functions \cite{Rao45,Cramer46}. QPET
gives the additional freedom to optimize over all possible
(POVM-)measurements \cite{Braunstein94} that generate the probability
distributions $p( \theta,A)$. Performing that optimization, one finds that
the standard deviation $\delta\theta$ of the fluctuations of $\theta$
estimated from $Q$ measurements are bounded from below by
the quantum Cram\'er-Rao bound (QCRB)
\begin{equation} \label{dl}
\delta \theta\ge\delta \theta_{\rm min}\equiv
\frac{1}{\sqrt{Q I(\hat{\rho}(\theta))}}\,,  
\end{equation}
 where $I(\hat{\rho}(\theta))=\sqrt{2d_{\rm Bures}(\hat{\rho}(\theta),\hat{\rho}(\theta+d\theta))}$
 is the Bures distance 
 between $\hat{\rho}(\theta)$ and $\hat{\rho}(\theta+d\theta)$ (also called quantum Fisher
 information), 
 defined as $d_{\rm 
 Bures}(\hat{\rho}_1,\hat{\rho}_2)=\sqrt{2}\sqrt{1-\sqrt{F(\hat{\rho}_1,\hat{\rho}_2)}}$ through the
 fidelity
 $F(\hat{\rho}_1,\hat{\rho}_2)=\tr(\hat{\rho}_1^{1/2}\hat{\rho}_2\hat{\rho}_1^{1/2})$.  
For pure states the fidelity reduces to the squared overlap of the two
 states,
$F(|\psi\rangle\langle\psi|,|\phi\rangle\langle\phi|)=|\langle\psi|\phi\rangle|^2$.\\

In \cite{Pinel12} we gave the formula for the Fisher information for an
arbitrary pure state described in terms of its Wigner function. 
With Eq.~(\ref{Wpm}), we therefore have the general translation of the change of
sensitivity from any pure state, whose Wigner function we know, to a state where a
single photon is subtracted or added in mode $k$.  In terms of the Fisher
information, 
\begin{equation} \label{dtmin}
I(\hat{\rho}^{(\pm,k)}(\theta))=2(2\pi)^M\int_{-\infty}^\infty
\left(W(\bX)'[\hat{\rho}^{(\pm,k)}]\right)^2d\bX^{2M}\,,
\end{equation}
where the $'$ means differentiation
with respect to $\theta$, the integral over $d\bX=dx_1\,dp_1\ldots
dx_M\,dp_M$ is over all modes, and we have corrected for a factor 2 due to a
different convention for the quadratures in \cite{Pinel12}. Note that $N^\pm$ will, in
general, also depend on $\theta$.

\section{QCR for single-photon-subtracted Gaussian states}
In the following we restrict ourselves to the single mode case and therefore
drop the index $k$. We assume that the co-variance
  matrix takes a diagonal form and we write $\bGamma={\rm 
  diag}(\Gamma_{xx},\Gamma_{pp})$ with $(\Gamma_{xx})^{-1}=2\langle (\hat{x}-\langle \hat{x}\rangle)^2\rangle$ and $(\Gamma_{pp})^{-1}=2\langle (\hat{p}-\langle \hat{p}\rangle)^2\rangle$.  This can always be
achieved by choosing an appropriate linear combination of the quadratures
$x$ and $p$. We do not consider the case where the rotation depends on the
parameter $\theta$. Introducing
  the 
$\theta$ dependence, the Wigner function of a Gaussian state is then given by
\begin{equation} \label{}
W_\theta(x,p)[\hat{\rho}]=\frac{\sqrt{\Gamma_{xx,\theta}\Gamma_{pp,\theta}}}{\pi}e^{-(x-\xb_\theta)^2\Gamma_{xx,\theta}-(p-\pb_\theta)^2\Gamma_{pp,\theta}}\,,  
\end{equation}
where the subscripts $\theta$ indicate the parameter dependence of the average
values of the quadratures and of the inverse covariance
matrix. We specialize on pure states for all values of $\theta$, in which
case one has $\Gamma_{xx,\theta}\Gamma_{pp,\theta}=1$.
Thus, we
are led to the final form of the Wigner function of a pure Gaussian
single-mode state
\begin{equation} \label{Gausspure}
W_\theta(x,p)[\hat{\rho}]=\frac{1}{\pi}e^{-(x-\xb_\theta)^2\Gamma_{\theta}-(p-\pb_\theta)^2/\Gamma_{\theta}}\,,  
\end{equation}
where we have abbreviated $\Gamma_{\theta}\equiv
\Gamma_{xx,\theta}$. For $\Gamma_\theta=1$, this is a coherent state,
otherwise a pure squeezed state. For reference below we note the Fisher
information for a pure Gaussian state with Wigner function
given by Eq.~\eqref{Gausspure} \cite{Pinel12} 
\begin{eqnarray}
I_G&=&\frac{4\Gamma_\theta \pb_\theta'^2+4\Gamma_\theta^3
  \xb_\theta'^2+\Gamma_\theta'^2}{2
  \Gamma_\theta^2}\,.  \label{eq:IGauss}
\end{eqnarray}
For a coherent state, this reduces to 
\begin{equation}
  \label{eq:Icoher}
  I_{\rm
    coher}=2(\xb_\theta'^2+\pb_\theta'^2) \,.
\end{equation}

\subsection{General result}
\begin{figure}[h]
\includegraphics[width=90mm]{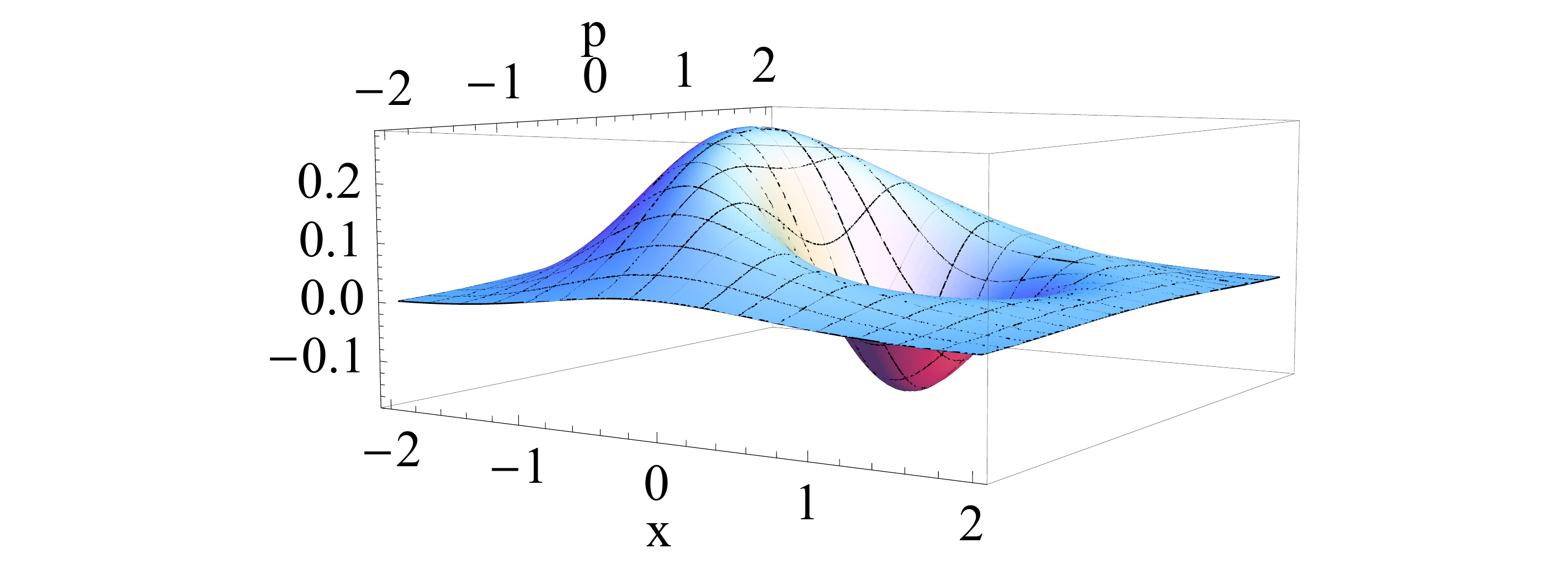}
\caption{Wigner function of photon-subtracted state for
  $\Gamma_\theta=1.1$, $\xb_\theta=1/20$.} \label{photon_sub_wigner}          
\end{figure}

Inserting Eq.~(\ref{Gausspure}) into Eq.~(\ref{Wpm}) we find the explicit Wigner
function of the 
photon-subtracted single-mode Gaussian state
\begin{eqnarray} \label{Gauss-1}
W_\theta(x,p)[\hat{\rho}^-]&=&\frac{e^{-\frac{(p-\pb_\theta)^2}{\Gamma_\theta}-(x-\xb_\theta)^2 \Gamma_\theta}}{\pi  \Gamma_\theta^2 \left(2 \left(-1+\pb_\theta^2+\xb_\theta^2\right)+\frac{1}{\Gamma_\theta}+\Gamma_\theta\right)} \big(2 p^2+2
\pb_\theta^2+4 p \pb_\theta (-1+\Gamma_\theta)\\
&&-\left(1+4 p^2\right) \Gamma_\theta+2 \left(1+p^2+x^2\right)
\Gamma_\theta^2+\left(-1-4 x^2+4 x \xb_\theta\right)
\Gamma_\theta^3+2 (x-\xb_\theta)^2 \Gamma_\theta^4\big)\,.\nonumber
\end{eqnarray}
The Wigner-function for a photon-subtracted state is
plotted in Fig.~\ref{photon_sub_wigner}.  We see that photon
subtraction leads to a large ``hole'' where the Wigner-function
becomes negative, confirming the strongly non-classical character of
such a state.
One checks that for $\Gamma_\theta=1$, Eq.~(\ref{Gauss-1})
gives back Eq.~(\ref{Gausspure}), confirming the
invariance of a coherent state under photon-subtraction.

The squared term in Eq.~\eqref{dtmin} can still be evaluated
analytically. We find that the Fisher information is given by
\begin{eqnarray}\label{eq:ISub}
I(\hat{\rho}^{(-)}(\theta))&=&
\frac{1}{2
  \left(\Gamma_\theta+2
  \left(-1+\pb_\theta^2+\xb_\theta ^2\right)
  \Gamma_\theta^2+\Gamma_\theta^3\right)^2}\nonumber\\
&&\times\Big\{4 \Gamma_\theta (3+\Gamma_\theta (-8+8
\pb_\theta^2+10 \Gamma_\theta\nonumber\\
&&+4 
\left(-2+\pb_\theta^2+\xb_\theta^2\right)
\left(\pb_\theta^2+\xb_\theta^2\right)
\Gamma_\theta+8 
\left(-1+\xb_\theta^2\right) \Gamma_\theta^2+3 
\Gamma_\theta^3))
\left(\pb_\theta'^2+\Gamma_\theta^2
\xb_\theta'^2\right)\nonumber\\
&&+32 \Gamma_\theta^2
(\pb_\theta
\left(1+\left(-1+\pb_\theta^2+\xb_\theta^2\right)
\Gamma_\theta\right) \pb_\theta'
-\xb_\theta
\Gamma_\theta
\left(-1+\pb_\theta^2+\xb_\theta^2+\Gamma_\theta\right)
\xb_\theta')\Gamma_\theta'\nonumber\\
&&+\Big(3+\Gamma_\theta  \Big(12
\pb_\theta^2+4
\left(-2+\pb_\theta^2+\xb_\theta^2\right)
\left(\pb_\theta^2+\xb_\theta^2\right)
\Gamma_\theta+12
\left(-1+\pb_\theta^2+\xb_\theta^2\right)
\Gamma_\theta^2+3 \Gamma_\theta^3\nonumber\\
&&+6 
\left(-2+2 \xb_\theta^2+3
\Gamma_\theta\right)\Big)\Big)\Gamma_\theta'^2\Big\} \label{sminex}
\end{eqnarray}
For $\xb'=\pb'=\Gamma_\theta'=0$, we find that
$I(\hat{\rho}^{(-)}(\theta))=0$, as it should be.
For $\Gamma_\theta=1\,\forall\theta$, Eq.~(\ref{eq:ISub}) simplifies greatly, and one recovers the result of Eq.~\eqref{eq:Icoher} for a coherent 
state.   But it is clear that in general the complexity of
Eq.~(\ref{eq:ISub}) 
cannot be avoided: differentiating $W(\bX)[\hat{\rho}^{\pm}]$ with respect to
$\theta$, together 
with the already present $x^2$ and $p^2$, leads to terms up to power 4 in $x$ and
$p$. Squaring the result, we obtain a
polynomial of order 8 in $x$ and $p$ as prefactor of the Gaussian, such that
the subsequent integration results in a corresponding polynomial containing
the elements 
of $\Gamma_\theta$, $\xb_\theta$ and $\pb_\theta$, as well as
their derivatives with respect to $\theta$. 

For the Gaussian state underlying the definition of the mean quadratures, 
$\xb_\theta$ and $\pb_\theta$ scale as $ \sim \sqrt{N}$.
This also leads to $\xb_\theta'\sim \sqrt{N}$ and
$\pb_\theta'\sim \sqrt{N}$, while we assume 
that 
$\Gamma_{\theta}$ is independent of $N$. 
These scalings allow us, for several special cases, to simplify the expression of $I(\hat{\rho}^{(-)}(\theta))$ at first orders and analyze its behavior.

\subsection{Special and limiting cases}\label{sec.spe}

We study in this subsection the asymptotic behavior of $I(\hat{\rho}^{(-)}(\theta))$ for several cases. \\

First, in the limit of large $\xb_\theta$, we have the asymptotic
expansion 
\begin{eqnarray}
I(\hat{\rho}^{(-)}(\theta))&=&\frac{4\Gamma_\theta \pb_\theta'^2+4\Gamma_\theta^3
  \xb_\theta'^2+\Gamma_\theta'^2}{2
  \Gamma_\theta^2}-\frac{4\xb_\theta' \Gamma_\theta'}{
  \Gamma_\theta
  \xb_\theta}+{\cal
  O}\left[\frac{1}{\xb_\theta}\right]^2\,.\label{Iasy} 
\end{eqnarray}
We recognize in the first term the result of Eq.~\eqref{eq:IGauss} for a Gaussian
state that 
scales as $N$, assuming that $\Gamma_\theta$ and at least one of $\xb'^2$ and
$\pb'^2$ are different from zero, such
that the numerator of the first term in Eq.~\eqref{Iasy} scales as 
$N$ for large $N$. Note that in \cite{Pinel12} the terms
with $\xb_\theta'^2$ have to be multiplied with a factor $1/2$ to
compare with the present result due to the different quadrature
convention. The second term will typically be of 
order $N^0$, as the scalings from $\xb_\theta$ and
$\xb_\theta'$  cancel under the same assumption. Thus
the result is only modified by a term of relative order $1/N$ compared to Gaussian states,
which is what one might have expected from the fact that one out of $N$
photons is 
taken out. In particular, the prefactor of the leading term $\propto N$ is
identical to the one of the squeezed Gaussian state, such that
asymptotically, photon 
subtraction does not enhance the sensitivity achievable with given squeezing
resources. \\  

Secondly, one checks that, for $\Gamma_\theta=1$, Eq.~\eqref{eq:ISub}
is invariant under the 
exchange of $\xb_\theta$ and $\pb_\theta$. In order to
simplify the analysis we will therefore set in the following $
\pb_\theta= \pb_\theta'=0$ for all $\theta$. 

\begin{figure}[h]
\includegraphics[width=90mm]{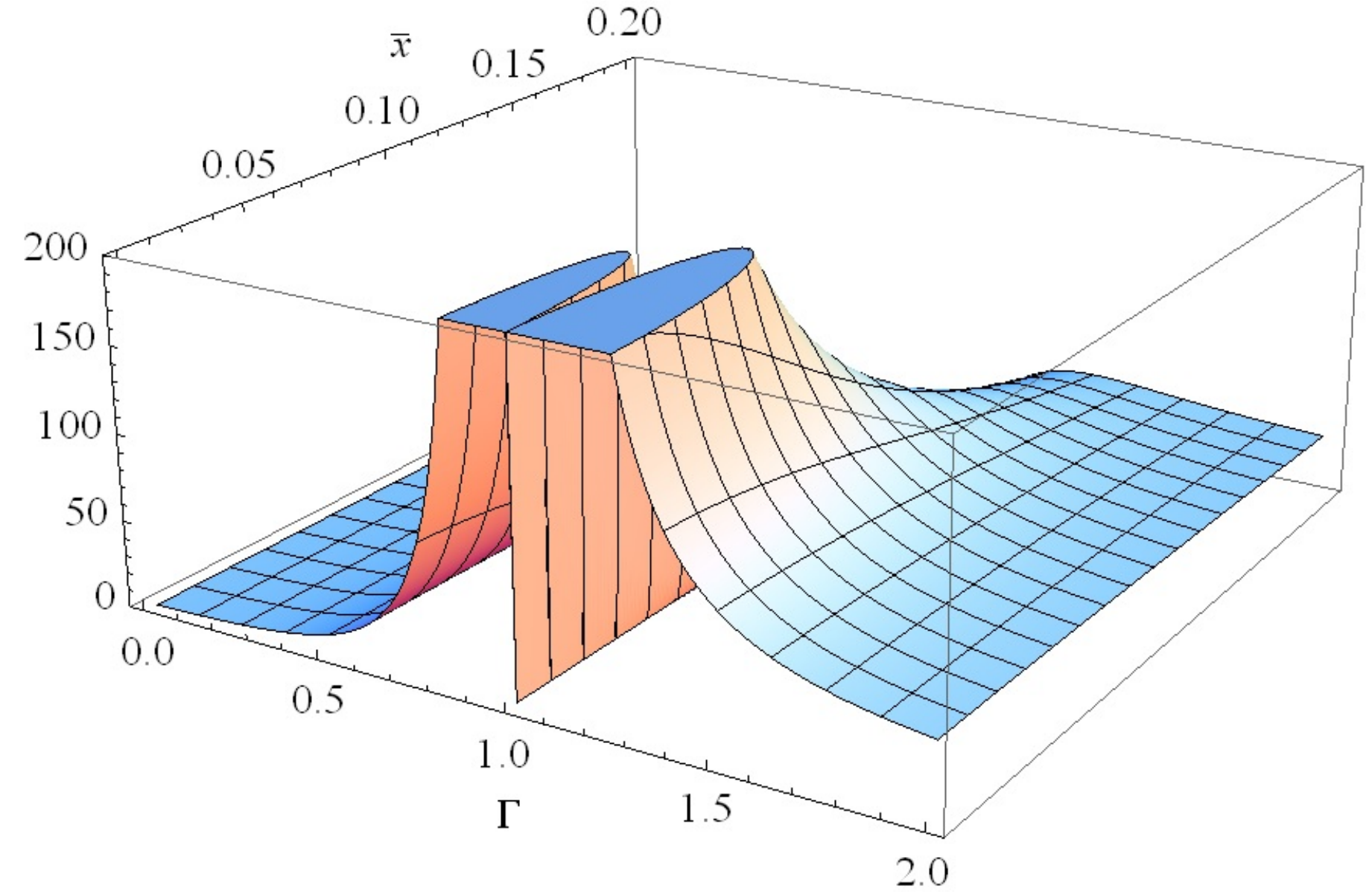}
\caption{Fisher information $I(\hat{\rho}(\theta))$ as function of
  $\xb\equiv \overline{x}_\theta$ and $\Gamma\equiv \Gamma_\theta$ for
  $\Gamma'_\theta=\pb_\theta=0$ in units of its value $2\pi$ for a
  coherent state ($\Gamma_\theta=1$) with the same $\xb$. The plot is
  cut at $I(\hat{\rho}(\theta))=200.$
}\label{qfi_photon_sub}         
\end{figure}

If all parameter dependence is in the shift in $x$-direction,
$\Gamma_\theta'=\pb_\theta'=\pb_\theta=0$, we have  
\begin{eqnarray}
I(\hat{\rho}^{(-)}(\theta))&=&
2\frac{\Gamma_\theta \left(3-8 \Gamma_\theta+2 \left(5-4
  \xb_\theta^2+2 \xb_\theta^4\right) \Gamma_\theta^2+8 \left(-1+\xb_\theta^2\right) \Gamma_\theta^3+3 \Gamma_\theta^4\right) \xb_\theta'^2}{ \left(1+2 \left(-1+\xb_\theta^2\right)
  \Gamma_\theta+\Gamma_\theta^2\right)^2}\label{gp0p0}
\,.
\end{eqnarray}
We see once more that this equation scales as $\sim N$, i.e.~for large $N$ one
cannot do much 
better than with the original Gaussian state.  However, for small $N$
($\xb_\theta\to 0$) and $\Gamma_\theta$ close to 1,
one gets an interesting divergence (see Fig.~\ref{qfi_photon_sub}). This can be attributed to the fact that the denominator vanishes for
$(\xb_\theta,\Gamma_\theta)=(0,1)$, which is the only root of the
denominator in the real plane.  Exactly at $\Gamma_\theta=1$ the numerator also vanishes and one gets of course back the finite result for the coherent state given by Eq.~(\ref{eq:Icoher}). 

When expanding Eq.~(\ref{gp0p0}) close to this point ($\Gamma_\theta=1+\epsilon_\theta$
with $|\epsilon_\theta|\ll 1$), we 
find
\begin{eqnarray}
I(\hat{\rho}^{(-)}(\theta))/(2\xb_\theta'^2)&=&
1+\left(1+\frac{2}{\xb_\theta^2}\right) \epsilon_\theta
+\left(\frac{1}{\xb_\theta^4}+\frac{1}{\xb_\theta^2}\right) \epsilon_\theta^2+{\cal
  O}(\epsilon_\theta )^3\label{gp0p02} 
\,.
\end{eqnarray}
Eq.~(\ref{gp0p02}) shows that indeed at $\epsilon_\theta=0$ we have the Fisher information of a coherent state, but at all finite values of $\epsilon_\theta$ one can reach arbitrarily large sensitivity in the limit of
$\xb_\theta\to 0$, as the lowest order in $\epsilon_\theta$ already 
diverges.  Furthermore, the two limits $\xb_\theta\to 0$ and
$\Gamma_\theta\to 1$ do not commute.  We just saw that taking
$\Gamma_\theta\to 1$ first gives the finite coherent state result also in
the limit $\xb_\theta\to 0$ (i.e. the Fisher information of the vacuum
state): $I(\hat{\rho}^{(-)}(\theta))=2\xb_\theta'^2$. However, the opposite order
of limits gives, at $\xb_\theta=0$, 
$I(\hat{\rho}^{(-)}(\theta))/\xb_\theta'^2=2\Gamma_\theta(3-2\Gamma_\theta+3\Gamma_\theta^2)/(\Gamma_\theta-1)^2$,
which diverges for $\Gamma_\theta\to 1$ as
\begin{equation}
  \label{eq:IvacG}
  I(\hat{\rho}^{(-)}(\theta))/\xb_\theta'^2=\frac{8}{(\Gamma_\theta-1)^2}+\frac{16}{\Gamma_\theta-1}+14+{\cal
    O}(\Gamma_\theta-1)\,. 
\end{equation}
Thus, the Fisher information is highly singular in the point
$(\xb_\theta,\Gamma_\theta)=(0,1)$, with a finite value on the line
$\Gamma_\theta=1$ when approaching $\xb_\theta\to 0$, but diverging on the
line $\xb_\theta=0$ when approaching $\Gamma_\theta\to 1$. Compared
with the Fisher information for a Gaussian state,
Eq.~(\ref{Gausspure}), that gives $2\Gamma_\theta \xb_\theta'^2$, we see
that subtracting a photon can greatly enhance the Fisher information
for the measurement of the same parameter.  

One may wonder how this
is compatible with the understanding that the quantum Cram\'er-Rao
bound for the 
Gaussian state \cite{Pinel12} gives the best possible sensitivity no
matter what POVM measurement is performed on the state, and no
matter how the data is analyzed.  In particular one might argue that
photon subtraction is achieved through interaction with another
physical system and subsequent measurement, which one might think is
describable by a set of POVMs.  The resolution of the apparent
paradox is through the observation that an essential step of photon
subtraction is the selection of a sub-ensemble, heralded by the
detection of a single photon as described above.  However, that
selection process makes the final state a non-linear function of the
initial density matrix and therefore cannot be described by
processing with a set
of POVMs summing up to the identity matrix. Thus, the previously
derived quantum Cram\'er-Rao bound \cite{Pinel12} does not apply
here, or in other words, photon subtraction (or addition) allows one
to escape from the limitations on state processing on which the
quantum Cram\'er-Rao bound is based. \\

We now study how useful the diverging Fisher information is.  In
particular, for
$\xb_\theta\to 0$ and in the limit of zero squeezing, the
preparation of the state as described in
\cite{wenger_non-gaussian_2004,kim_recent_2008} by  
post-selection heralded on a single detected photon after passing through
the beam splitter will fail almost always, such
that the total measurement time including state preparation increases and the experimentally relevant
sensitivity per square root of Hertz is reduced. Therefore, when taking the preparation time
into account, the quantum Fisher information has to be
appropriately rescaled, and the question is whether this removes its
divergence. 

A first observation is that the preparation scheme by
\cite{wenger_non-gaussian_2004,kim_recent_2008} is by no way unique.  There
might be more efficient preparation schemes that require a different
renormalization of the Fisher information, or maybe none at all.
Nevertheless, it is instructive to calculate the required renormalization
for the particular preparation scheme in
\cite{wenger_non-gaussian_2004,kim_recent_2008}. As we will show in the following, it turns out that the
divergence of the Fisher information is completely
removed. Therefore, when taking into account the
increasing preparation time of the photon-subtracted state in the limit of
an initial unsqueezed vacuum state, the experimentally relevant sensitivity
per square root of Hertz cannot be increased to
arbitrarily high levels, at least with this preparation scheme.

To demonstrate this, let us calculate the success probability for the
preparation scheme, i.e.~the probability to detect exactly one photon of the
initial squeezed coherent state in the darker of the two output ports after
it passes an almost transparent beam   
splitter. The two mode unitary transformation
that describes the beam splitter is given by
$\hat{U}_{BS}(\delta)=\exp(\delta(\hat{a}^\dagger \hat{b}-\hat{a} \hat{b}^\dagger))$, where $\hat{a},\hat{b}$ are
annihilation operators respectively for the two modes, and $\delta$ is the
mixing angle. Owing to the conservation of total photon-number $N$ by the
beam splitter, 
it is convenient to represent $\hat{U}_{BS}$ in the dual-rail basis,
\begin{eqnarray}
  \langle k, N-k |\hat{U}_{BS}|m,
  N-m\rangle&=&\sqrt{\frac{k!(N-k)!}{m!(N-m)!}}\sum_{l={\rm Max}(m-k,0)}^{{\rm
      Min}(N-k,m)} {m \choose l}{N-m \choose N-k-l}\nonumber\\
&&\times (-1)^l\left(\cos\delta\right)^{m+N-k-2l}\left(\sin\delta\right)^{k-m+2l} \label{eq:BS}\,,
\end{eqnarray}
where the kets $|n,m\rangle$ denotes a product of photon number eigenstates
with $n$ and $m$ photons, respectively, in the two modes
\cite{Nielsen00,Braun06}. 

Next we express the initial state $|\psi\rangle$
very generally in the photon number basis as $|\psi\rangle=\sum_{n=0}^\infty
a_n|0,n\rangle$, with the first  mode initially in the vacuum state.  After
the action of the beam splitter we have the final state
$|\psi'\rangle=\hat{U}_{BS}|\psi\rangle$.  The probability to detect one photon in
the first mode is then given by $P^{\mathrm{out}}_1=\sum_{n=0}^\infty |\langle
1,n|\psi'\rangle|^2$.   A few lines of calculation lead to
\begin{equation}
  \label{eq:p1gen}
  P^{\mathrm{out}}_1=\sum_{n=1}^\infty P_n n \left(\cos\delta\right)^{2(n-1)}\sin^2\delta\,,
\end{equation}
where $P_n=|a_n|^2$ denotes the initial probabilities for $n$ photons in the
input mode. For a squeezed coherent state, these probabilities are well known (see
e.g.~Eq.~(3.5.16) in \cite{Scully97}).  We adapt the notation of that
reference in writing the squeezed coherent state as
\begin{equation}
  \label{eq:squeeze}
  |\alpha,\xi\rangle=\hat{S}(\xi)\hat{D}(\alpha)|0\rangle\,,
\end{equation}
where $\hat{S}(\xi)$ and $\hat{D}(\alpha)$ are the usual squeeze and displacement
operators, $\hat{S}(\xi)=\exp((\xi^*\hat{a}^2-\xi (\hat{a}^\dagger)^2 )/2)$ with $\xi=r e^{i
  \vartheta}$, $(r\ge 0$) and 
$\hat{D}(\alpha)=\exp(\alpha \hat{a}^\dagger -\alpha^* \hat{a})$. Then
\begin{equation}
  \label{eq:pnsq}
  P_n=\frac{(\tanh(r))^n}{2^n n! \cosh(r)}\exp\left(-|\alpha|^2+\frac{1}{2}
\left(e^{-i\vartheta}\alpha^2 +e^{i\vartheta}(\alpha^*)^2\right)\tanh(r)
\right)|H_n\left( \frac{\alpha e^{-i\vartheta/2}}{\sqrt{\sinh(2r)}}\right)|^2\,,  
\end{equation}
where $H_n(x)$ is the Hermite polynomial of order $n$. 

\begin{figure}[h]
\includegraphics[width=90mm]{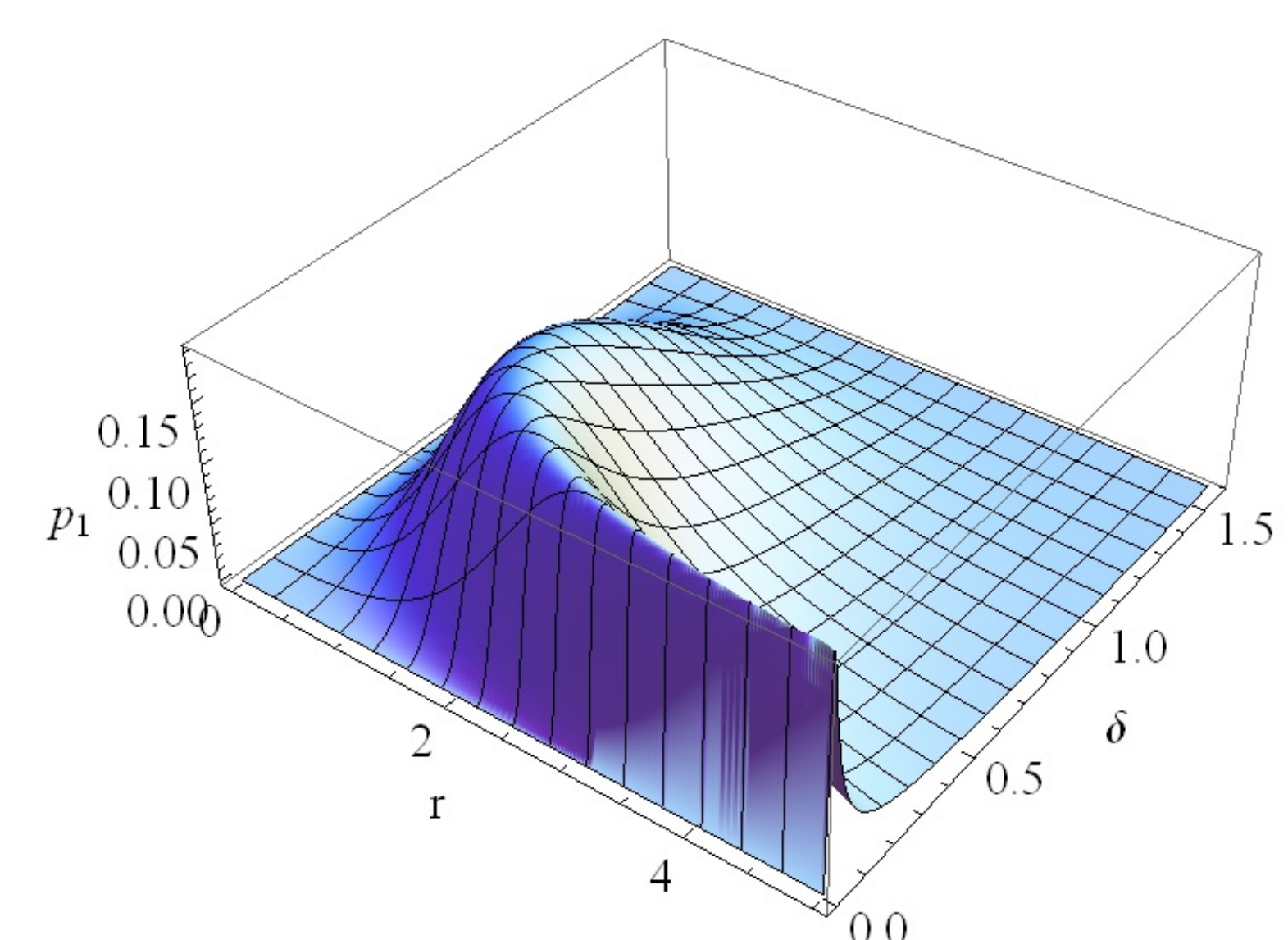}
\caption{Success probability $p_1\equiv P_1^{\rm out}$) for preparation
  of the single photon subtracted 
  squeezed coherent state for $\xb_\theta=0$ as function of squeeze
  parameter $r$ and beam splitter mixing angle $\delta$. }\label{fig.p1}          
\end{figure}

A closed form of
$P_n$ can be found in the case that interests us most, namely at
$\alpha=0$, where we find 
  \begin{equation}
    \label{eq:pnsq2}
    P^{\mathrm{out}}_1(\alpha=0,\delta)=\frac{\sin^2(2\delta)\sech(r)\tanh^2(r)}{4(1-\cos^4\delta\tanh^2r)^{3/2}}\,. 
  \end{equation}
The function is  $\pi$-periodic in $\delta$ as to be expected.  The
squeezing angle $\vartheta$ has disappeared with the amplitude $\alpha\to 0$,
as can be seen from Eq.~\eqref{eq:pnsq}. A
plot of $P^{\mathrm{out}}_1$ as function of the remaining parameters $r$ and $\delta$ is
shown in Fig.~\ref{fig.p1}. We see that $P^{\mathrm{out}}_1$, as function of
$\delta$, reaches a sharp maximum of about 0.2 if the squeezing is rather strong. The
maximum moves with increasing squeezing closer and closer to $\delta=0$.
For small squeezing $P^{\mathrm{out}}_1(\alpha=0,\delta)$ starts off quadratically, as is confirmed by
expanding it about $r=0$,
\begin{equation}
P^{\mathrm{out}}_1(\alpha=0,\delta)=\frac{1}{4}\sin^2(2\delta)r^2+{\cal O}(r^3)\mbox{ for }
r\ll 1\,.   \label{eq:p1'}
\end{equation}

Next we need to calculate the parameters $\Gamma_\theta$ and
$x_\theta$ used so far for characterizing our Gaussian states.  This is
easily achieved with the results for the expectation values for
$\hat{a}$, $\hat{a}^2$ 
and $\hat{a}^\dagger \hat{a}$ for squeezed coherent states found in Eq.~(2.7.11) in
\cite{Scully97}. Inserting the result into the equation leading to
Eq.~\eqref{eq:IvacG}, we find for the 
Fisher information, in the case of
$\pb_\theta=\pb_\theta'=\Gamma'=0 $,
\begin{equation}
  \label{eq:gp0p03}
  I(\hat{\rho}^{(-)}(\theta))=\xb_\theta'^2\frac{4e^{4r}(-1+3\cosh(2r))}{(-1+e^{2r})^2}\,,
\end{equation}
which for small $r$ diverges as $1/r^2$,
\begin{equation}
  \label{eq:gp0p04}
  I(\hat{\rho}^{(-)}(\theta))=2\xb_\theta'^2\left(\frac{1}{r^2}+\frac{2}{r}+\frac{14}{3}+{\cal
    O}(r^1)\right)
\end{equation}
in agreement with Eq.~(\ref{eq:IvacG}).
We are interested in the regime where $P^{\mathrm{out}}_1$ is  small. This means that on
the average one has to repeat the preparation attempt of the
photon-subtracted state a number of times that scales as $1/P^{\mathrm{out}}_1$. Thus, in a
given finite time, the number of measurements that can be done with a
successfully prepared photon-subtracted state is proportional to
$P^{\mathrm{out}}_1$. (In other words, if a single preparation takes a time $T$, the
preparation rate is $P^{\mathrm{out}}_1/T$). Since the quantum Fisher information for a single shot measurement given by Eq.~(\ref{dl}) is scaled by the number of measurements $Q$ in the QCRB, we have to rescale the 
QFI with a factor $P^{\mathrm{out}}_1$ in order to find the scaling of the
effective quantum Fisher information that includes
the preparation rate (giving a sensitivity with units
$[\theta]\sqrt{\rm Hz}$). We see from Eq.~\eqref{eq:p1'} that the scaling of 
$P^{\mathrm{out}}_1$ with $r$ 
at small $r$ is quadratic, so it exactly cancels the
$1/r^2$ divergence of the quantum Fisher information. The effective quantum Fisher information reads
\begin{equation}
  \label{eq:pI}
 I_\mathrm{eff}(\hat{\rho}^{(-)}(\theta))= \frac{1}{
   2}\xb_\theta'^2\sin^2(2\delta)+{\cal O}(r) \,.
\end{equation}
%

Let us stress the parallel between this result and noiseless linear amplification, as first proposed by Xiang et al. in \cite{xiang_heralded_2010}. In the same way as the noiseless amplification of an initial quantum state can occur non-deterministically by photon heralding, in the case we have studied it is possible to surpass the sensitivity of a Gaussian state in parameter estimation by photon subtraction, but only in a non deterministic way that is conditioned by the successful subtraction of one photon.

All of the above remains valid if the parameter dependence of the Gaussian
state is carried by $\pb_\theta$ instead of
$\xb_\theta$. The quantum Fisher information at $\alpha=0$ is then simply to be
multiplied with $1/\Gamma_\theta^2$, which does not change the behavior at
$\Gamma_\theta\to 1$.  \\

Finally, if all parameter dependence is in
$\Gamma_\theta$, i.e.~if both shifts are independent of $\theta$,
$\xb_\theta'=\pb_\theta'=\pb=0$, we get a result that is
asymptotically independent of $N$,
\begin{eqnarray}
 I(\hat{\rho}^{(-)}(\theta))&=&
\frac{\left(3+\Gamma_\theta \left(4 \xb^4 \Gamma_\theta+4
  \xb_\theta^2 \left(3-2 
  \Gamma_\theta +3 \Gamma_\theta^2\right)+3 \left(-4+6
  \Gamma_\theta-4 \Gamma_\theta^2+\Gamma_\theta^3\right)\right)\right) \Gamma_\theta'^2}{2  \left(\Gamma_\theta+2
  \left(-1+\xb_\theta^2\right) 
\Gamma_\theta^2+\Gamma_\theta^3\right)^2}
\end{eqnarray}
paralleling once more the behavior for Gaussian states \cite{Pinel12}. In
the limit of initial vacuum, $\xb_\theta=\pb_\theta=0$,
one finds $I(\hat{\rho}^{(-)}(\theta))=3\Gamma_\theta'^2/(2 \Gamma_\theta^2)$,
i.e.~there is no divergence at $\Gamma_\theta\to 1$. 


\section{QCR for Gaussian states with one photon added}
\subsection{General results}
Photon addition leads to  more complicated expressions, but the
procedure for obtaining the QCR follows the same pattern as above.  In order
to simplify expressions a little we drop all subscripts $\theta$ in this
section. 
We first find the Wigner function for the photon-added Gaussian state, 
\begin{eqnarray}
  \label{eq:Gauss+1}
W(x,p)[\hat{\rho}^+]&=&\frac{e^{-\frac{(p-\pb)^2}{\Gamma}-(x-\xb)^2
    \Gamma} }{\pi  \Gamma^2 \left(2
    \left(3+\pb^2+\xb^2\right)+\Gamma
    ^{-1}+\Gamma\right)}\\
&\times&\Big(2 p^2+2
\pb^2+\left(-1+4 p^2\right) \Gamma+2 \left(1+p^2+x^2\right) \Gamma^2+\left(-1+4 x^2-4
x \xb\right) \Gamma^3\nonumber\\
&&+2 (x-\xb)^2 \Gamma^4-4 p \pb (1+\Gamma
 )\Big)\nonumber\,.
\end{eqnarray}
Inserting this in Eq.~(\ref{dtmin}), expanding in $x$ and $p$ up to the tenth
order, integrating
symbolically term by 
term, and adding up the terms from the expansion, we find the exact
expression for the Fisher information, which can be found in the Appendix, in
Eq.~(\ref{eq:I+}).

\subsection{Special and limiting cases}
The expansion for large $\xb$ leads to the exact same expression for the
first two highest order terms in $N$ as for photon subtraction in Eq.~(\ref{Iasy}).  Thus, for large photon numbers, photon subtraction and
addition are essentially equivalent concerning their usefulness for
precision measurements, and, as mentioned above, the increase (or
decrease, depending on the sign of the second term in Eq.~(\ref{Iasy})) is of
relative order $1/N$ only. \\

If all the parameter dependence for a state centered at $\pb=0$ is in the
shift in $x$-direction, $\Gamma'=0=\pb=\pb'=0$, we find 
\begin{eqnarray}  \label{eq:I+p0G0}
I(\hat{\rho}^{(+)}(\theta))&=&
  \frac{2}{(1+2
      (3+\xb^2) \Gamma+\Gamma^2)^4}\\
&\times & \Big\{\Gamma(3+4 (13+3 \xb^2)
      \Gamma+4 (83+8 \xb^2+4 \xb^4)
      \Gamma^2+4 (239+135 \xb^2+52 \xb^4+4
        \xb^6) \Gamma^3\nonumber\\
&&+2 (593+784 \xb^2+432
        \xb^4+96 \xb^6+8 \xb^8) \Gamma^4+4
      (91+177 \xb^2+84 \xb^4+12 \xb^6)
      \Gamma^5\nonumber\\
&&+4 (35+48 \xb^2+12 \xb^4)
      \Gamma^6+4 (9+5 \Gamma^2) \Gamma^7+3
      \Gamma^8) \xb'^2
\Big\}\,.\nonumber
\end{eqnarray}
We see once more the leading behavior $\propto N$ due to the terms
$\xb^8\xb'^2/\xb^8$. However, contrary to photon subtraction, no divergence is observed for $N\ll 1$, regardless of the squeezing, as the
possibly vanishing term $(-1+\xb^2)$ in the denominator for the photon-subtracted case is now replaced by $3+\xb^2>0$. \\

Finally, if both shifts are independent of $\theta$,
$\xb'=\pb'=\pb=0$, we have
\begin{eqnarray}
  \label{eq:I+x0p0}
I(\hat{\rho}^{(+)}(\theta))&=&  \frac{1}{2\Gamma^2 (1+2 (3+\xb^2)
  \Gamma+\Gamma^2)^4}\\
&\times &
\Big\{
(3+24 (2+\xb^2) \Gamma+4 (105+88 \xb^2+16 \xb^4)
\Gamma^2+8 (114+149 \xb^2+60 \xb^4+8 \xb^6) \Gamma^3\nonumber\\
&&+2 (665+992 \xb^2+480
  \xb^4+96 \xb^6+8 \xb^8) \Gamma^4+8 (114+149 \xb^2+60 \xb^4+8 \xb^6)
  \Gamma^5\nonumber\\
&&+4 (105+88 \xb^2+16 \xb^4) \Gamma^6+24 (2+\xb^2) \Gamma^7+3
  \Gamma^8) (\Gamma')^2\Big\}\,.  \nonumber
\end{eqnarray}

In the limit of initial vacuum (i.e.~in addition $\xb\to 0$), the expression converges to 
\begin{eqnarray}
  \label{eq:I+x0p01}
I(\hat{\rho}^{(+)}(\theta))&=&
  \frac{\left(3+48 \Gamma+420 \Gamma^2+912 \Gamma^3+1330 \Gamma^4+912 \Gamma^5+420 \Gamma^6+48 \Gamma^7+3 \Gamma^8\right)
(\Gamma')^2}{ 2\Gamma^2 \left(1+6 \Gamma+\Gamma^2\right)^4} \,.
\end{eqnarray}
For large $\Gamma$ this expression decays just as in the photon subtracted state, i.e.~as $3(\Gamma')^2/(2\Gamma^2)$.

\section{Conclusions}

For large number of photon $N$, subtraction (or addition) of a single photon from (or to) a pure Gaussian state does not
substantially alter the scaling with $N$ of the sensitivity with which one can estimate a
parameter $\theta$ coded in the initial Gaussian state.  The
corrections to the quantum Fisher information are only of relative order
$1/N$. 
For small $N$, photon subtraction can increase
the sensitivity attainable with squeezed states, in particular for
almost vanishing squeezing parameter $r$ and $N\ll 1$. The quantum Fisher
information diverges as $1/r^2$ in that limit, reflecting the extremely
non-classical behavior of such a state.  However, in the standard
preparation scheme, 
based on the passage of a squeezed coherent state through a beam splitter that is almost
transparent for the state, and heralding an output based on the detection of
a single photon in the almost dark output port
\cite{wenger_non-gaussian_2004,kim_recent_2008}, the success probability
of the preparation decays proportionally to $r^2$ with the squeezing
parameter.  The 
rescaled quantum Fisher information for the experimentally relevant
sensitivity in a fixed bandwidth that takes into 
account the preparation time of the state is given by the product of
the success probability and the single shot quantum Fisher 
information.  This leads to an exact cancellation of the
divergence of the 
Fisher information.  It remains to be seen whether there are
deterministic preparation schemes or experimental niches where such
states that 
use essentially no light at all can compete with the standard approach of
very large photon numbers. 

{\em Acknowledgments: } We thank Claude Fabre for useful
discussions. O.P.~acknowledges support by the Australian Research
Council Centre of Excellence for Quantum Computation and Communication
Technology, project number CE110001027. This work is supported by the
European Research Council starting grant Frecquam. 

\section{Appendix}
Here we report the exact Fisher information for the photon-added state.  For
improving the readability, we skip all the subscripts $\theta$, but it is
understood that $\xb$, $\pb$, and $\Gamma$ depend on $\theta$, and $'$
denotes $d/d\theta$. 
\begin{eqnarray} \label{eq:I+}
 I(\hat{\rho}^{(+)}(\theta))&=&\frac{1}{2  \Gamma ^2
(1+2 (3+\pb^2+\xb^2) \Gamma +\Gamma^2)^4}\\
&\times&\Big\{
16 (9+3 \pb^2+5 \xb^2) \Gamma ^{10} (\xb')^2+12 \Gamma ^{11} (\xb')^2+4
\Gamma^9 
(3 (\pb')^2+4 (35+4 \pb^4+12 \xb^2 (4+\xb^2)\nonumber\\
&&+8 \pb^2 (3+2 \xb^2))
(\xb')^2)+3 (\Gamma ')^2+\Gamma ^8 (16 (13+3 \pb^2+5 \xb^2) (\pb')^2+16
(91+103 \pb^2\nonumber\\
&&+36 \pb^4+4 \pb^6+(177+20 \pb^2 (6+\pb^2)) \xb^2+28 (3+\pb^2)
\xb^4+12 \xb^6) (\xb')^2-32 \xb \xb' \Gamma '\nonumber\\ 
&&+3 (\Gamma ')^2)+12 \Gamma
((\pb')^2+2
(2+\pb^2+\xb^2) (\Gamma ')^2)+4 \Gamma ^2 (4 (9+5 \pb^2+3 \xb^2) (\pb')^2+8
\pb \pb' \Gamma '\nonumber\\  
&&+(7+4 \pb^2+4 \xb^2) (15+4 \pb^2+4 \xb^2)
(\Gamma')^2)+2
\Gamma ^4 (8 (91+12 \pb^6+103 \xb^2+28 \pb^4 (3+\xb^2)\nonumber\\
&&+4 \xb^4 (9+\xb^2)+\pb^2
(177+20 \xb^2 (6+\xb^2))) (\pb')^2+8 (13+5 \pb^2+3 \xb^2) (\xb')^2\nonumber\\
&&+16
(2 \pb (33+20 \xb^2+4 (\pb^4+\xb^4+\pb^2 (5+2 \xb^2))) \pb'\nonumber\\
&&-\xb
(45+4 \pb^4+8 \pb^2 (4+\xb^2)+4 \xb^2 (8+\xb^2)) \xb') \Gamma '+(665+992
\xb^2+8 (\pb^8+4 \pb^6 (3+\xb^2)\nonumber\\
&&+6 \pb^4 (10+6 \xb^2+\xb^4)+\xb^4 (60+12
\xb^2+\xb^4)+4 \pb^2 (31+30 \xb^2+9 \xb^4+\xb^6))) (\Gamma ')^2)\nonumber\\
&&+8
\Gamma ^7 (2 (83+8 \pb^2+4 \pb^4+16 (4+\pb^2) \xb^2+12 \xb^4) (\pb')^2\nonumber\\
&&+(593+784
\xb^2+8 (\pb^8+4 \pb^6 (3+\xb^2)+6 \pb^4 (3+\xb^2)^2+\xb^4 (54+12
\xb^2+\xb^4)\nonumber\\
&&+2 
\pb^2 (53+2 \xb^2 (27+9 \xb^2+\xb^4)))) (\xb')^2\nonumber\\
&&-4 \xb (19+5
\pb^2+5 \xb^2) \xb' \Gamma '+3 (2+\pb^2+\xb^2) (\Gamma ')^2+4 \pb \pb' (-32
\xb \xb'+(-1+\pb^2+\xb^2) \Gamma '))\nonumber\\
&&+8 \Gamma ^5 ((593+848 \xb^2+8 (\pb^8+4
\pb^6 (3+\xb^2)+6 \pb^4 (3+\xb^2)^2+\xb^4 (54+12 \xb^2+\xb^4)\nonumber\\
&&+2 \pb^2
(49+2 \xb^2 (27+9 \xb^2+\xb^4)))) (\pb')^2+2 (83+8 \xb^2+4 (3
\pb^4+\xb^4+4 \pb^2 (4+\xb^2))) (\xb')^2\nonumber\\
&&-8 \xb (3+\pb^2+\xb^2)
(21+2 \pb^4+4 \pb^2 (3+\xb^2)+2 \xb^2 (6+\xb^2)) \xb' \Gamma '\nonumber\\
&&+(114+8 \pb^6+149
\xb^2+60 \xb^4+8 \xb^6+12 \pb^4 (5+2 \xb^2)+\pb^2 (149+24 \xb^2 (5+\xb^2)))
(\Gamma ')^2\nonumber\\
&&+8 \pb \pb' (-16 \xb \xb'+(3+\pb^2+\xb^2) (21+2 \pb^4+4 \pb^2
(3+\xb^2)+2 \xb^2 (6+\xb^2)) \Gamma '))\nonumber\\
&&+4 \Gamma ^6 (4 (239+135 \pb^2+52
\pb^4+4 \pb^6+(273+152 \pb^2+20 \pb^4) \xb^2+4 (25+7 \pb^2) \xb^4+12 \xb^6)
(\pb')^2\nonumber\\
&&+4 (239+12 \pb^6+135 \xb^2+52 \xb^4+4 \xb^6+4 \pb^4 (25+7 \xb^2)+\pb^2
(273+152 \xb^2+20 \xb^4)) (\xb')^2\nonumber\\
&&-16 \xb (33+20 \xb^2+4 (\pb^4+\xb^4+\pb^2
(5+2 \xb^2))) \xb' \Gamma '\nonumber\\
&&+(7+4 \pb^2+4 \xb^2) (15+4 \pb^2+4 \xb^2)
(\Gamma ')^2+8 \pb \pb' (-64 \xb \xb'+(45+4 \pb^4+8 \pb^2 (4+\xb^2)\nonumber\\
&&+4 \xb^2
(8+\xb^2)) \Gamma '))+4 \Gamma ^3 (4 (35+12 \pb^4+16 \pb^2 (3+\xb^2)+4
\xb^2 (6+\xb^2)) (\pb')^2+3 (\xb')^2\nonumber\\
&&+8 \pb (19+5 \pb^2+5 \xb^2)
\pb' \Gamma '+2 \Gamma ' (-4 \xb (-1+\pb^2+\xb^2) \xb'\nonumber\\
&&+114 \Gamma '+(\pb^2+\xb^2)
(149+8 \pb^4+60 \xb^2+8 \xb^4+4 \pb^2 (15+4 \xb^2)) \Gamma '))\Big\}\,.\nonumber
\end{eqnarray}

\bibliography{phosub_biblio}

\end{document}